\documentclass[preprint,aps,showpacs]{revtex4}

\usepackage{graphicx}
\usepackage{textcomp}

\begin{document}
\draft
\title{Klein tunneling of light in fiber Bragg gratings}
  \normalsize

\author{Stefano Longhi}
\address{Dipartimento di Fisica, Politecnico di Milano, Piazza L. da Vinci 32, I-20133 Milano, Italy}

%\date{.}

%
\bigskip
\begin{abstract}
\noindent A photonic analogue of Klein tunneling (KT), i.e. of the
exotic property of relativistic electrons to pass a large repulsive
and sharp potential step, is proposed for pulse propagation in a
nonuniform fiber Bragg grating with an embedded chirped region. KT
can be simply observed as the opening of a transmission window
inside the grating stop band, provided that the impressed chirp is
realized over a length of the order of the analogue of the Compton
wavelength.
\end{abstract}

%\pacs{11.30.Er,42.50.Xa}

\maketitle

\section{Introduction}
A remarkable prediction of the Dirac equation is that a
below-barrier electron can pass a large repulsive and {\em sharp}
potential step without the exponential damping expected for a
non-relativistic particle. Such a transparency effect, originally
predicted by Klein \cite{Klein} and referred to as Klein tunneling
(KT), arises from the existence of negative-energy solutions of the
Dirac equation and requires a potential step height $\Delta V$ of
the order of twice the rest energy $mc^2$ of the electron
\cite{Calogeracos99}. Relativistic tunneling across a {\em smooth}
potential step, which describes the more physical situation of a
constant electric field $E$ in a finite region of space of length
$l$, was subsequently studied by Sauter \cite{Sauter}. Sauter showed
that to observe barrier transparency the potential increase $\Delta
V \simeq eEl$ should occur over a distance $l$ of the order or
smaller than the Compton wavelength $\lambda_C=\hbar/(mc)$, the
transmission probability rapidly decaying toward zero for a smoother
potential increase \cite{ Calogeracos99,Sauter,Emilio}. The required
field corresponds to the critical field for $e^+e^-$ pair production
in vacuum, and its value is extremely strong making the observation
of relativistic KT for electrons very challenging. Therefore,
growing efforts have been devoted to find experimentally accessible
systems to investigate analogs of relativistic KT \cite{NO}.
Recently, great interest has suscitated the proposal \cite{GR1} and
first experimental evidences \cite{GR2,GR3} of KT for
non-relativistic electrons in graphene, which behave like massless
Dirac fermions. On the other hand, optics has offered on many
occasions a test bed to investigate the dynamical aspects embodied
in a wide variety of coherent quantum phenomena (see, for instance,
\cite{Longhi09} and references therein). In optics, several
proposals of KT analogs have been suggested as well, including light
propagation in deformed honeycomb photonic lattices \cite{Segevun}
whose band structure is similar to the one of graphene
\cite{Seg1,Seg2}, light refraction at the interface between
positive-index and negative-index media \cite{meta}, spatial light
propagation in binary waveguide arrays \cite{Longhi10}, and
stationary light pulses in an atomic ensemble with
electromagnetically induced transparency \cite{PRL09EIT}. The
experimental implementations of such schemes, however, might be a
nontrivial matter, and an experimental observation of KT for photons
is still lacking. On the other hand, multilayer and Bragg dielectric
structures, such as fiber Bragg gratings (FBGs), are rather simple
photonic devices with flexible design that have been successfully
demonstrated to provide an accessible laboratory tool to investigate
photonic analogues of non-relativistic tunneling phenomena
\cite{Steinberg93,Longhi02,Longhi03}. Here it is shown that an
optical analogue of KT can be achieved in a nonuniform FBG composed
by two periodic sections linked by a chirped section which mimics an
external potential step in the Dirac equation. Such a FBG-based
system might be considered the simplest system proposed so far in
order to observe Klein tunneling in any optical system.

\section{Quantum-optical analogy}
The starting point of our analysis if provided by a standard model
of light propagation in a FBG with a longitudinal refractive index $
n(z')=n_0+\Delta n \; m(z') \cos[2 \pi z' / \Lambda +2 \phi(z')]$,
where $n_0$ is the effective mode index in absence of the grating,
$\Delta n \ll n_0$ is the peak index change of  the grating,
$\Lambda$ is the nominal period of the grating defining the
reference frequency $\omega_B=\pi c/(\Lambda n_0)$ of Bragg
scattering, $c$ is the speed of light in vacuum, and $m(z')$, $2
\phi(z')$ describe the slow variation, as compared to the scale of
$\Lambda$, of normalized amplitude and phase, respectively, of the
index modulation. Note that the local spatial frequency of the
grating is $k(z')=2 \pi / \Lambda+2(d \phi/dz')$, so that the local
chirp rate is $C=dk/dz'=2(d^2 \phi/dz'^2)$. The periodic index
modulation leads to Bragg scattering between two counterpropagating
waves at frequencies close to $\omega_B$. By letting
$E(z',t)=\varphi_1(z',t) \exp[-i \omega_B t +ik_B z'+i \phi(z')]+
\varphi_2(z',t) \exp[-i \omega_B t -ik_B z'-i \phi(z')]+c.c.$ for
the electric field in the fiber, where $k_B=\pi/\Lambda$, the
envelopes $\varphi_1$ and $\varphi_2$ of counterpropagating waves
satisfy the coupled-mode equations \cite{Sipe}
\begin{eqnarray}
i \left[ \partial_{z'} + (1/v_g) \partial_t
\right] \varphi_1 & = & (d \phi/dz') \varphi_1-\kappa(z') \varphi_2 \\
i \left[ -\partial_{z'} +(1/v_g) \partial_t  \right] \varphi_2 & = &
(d \phi/dz') \varphi_2-\kappa(z') \varphi_1
\end{eqnarray}
where $\kappa(z') \equiv [k_B m(z') \Delta n ]/(2n_0)$ and $v_g \sim
c/n_0$ is the group velocity at the Bragg frequency. The analogy
between pulse propagation in the FBG and the Dirac equation in
presence of an electrostatic field is at best captured by
introducing the dimensionless variables $z=z'/Z$ and $\tau=t/T$ ,
with characteristic spatial and time scales $Z=2 n_0/(k_B \Delta n)$
and $T=Z/v_g$, and the new envelopes $\psi_{1,2}(z')=[\varphi_1(z')
\mp \varphi_2(z')]/ \sqrt 2$. In this way, Eqs.(1-2) can be cast in
the Dirac form
\begin{equation}
i \partial_{\tau} \psi=-i \sigma_1 \partial_z \psi+m \sigma_3 \psi
+V(z) \psi
\end{equation}
for the spinor wave function $\psi=(\psi_1,\psi_2)^T$, where
$V(z)=(d \phi /dz)$ and $\sigma_{1,3}$ are the Pauli matrices,
defined by
\begin{equation}
\sigma_1= \left(
\begin{array}{cc}
0 & 1 \\
1 & 0
\end{array}
\right) \; , \; \; \sigma_3= \left(
\begin{array}{cc}
1 & 0 \\
0 & -1
\end{array}
\right).
\end{equation}

\begin{figure}[htb]
\centerline{\includegraphics[width=8.2cm]{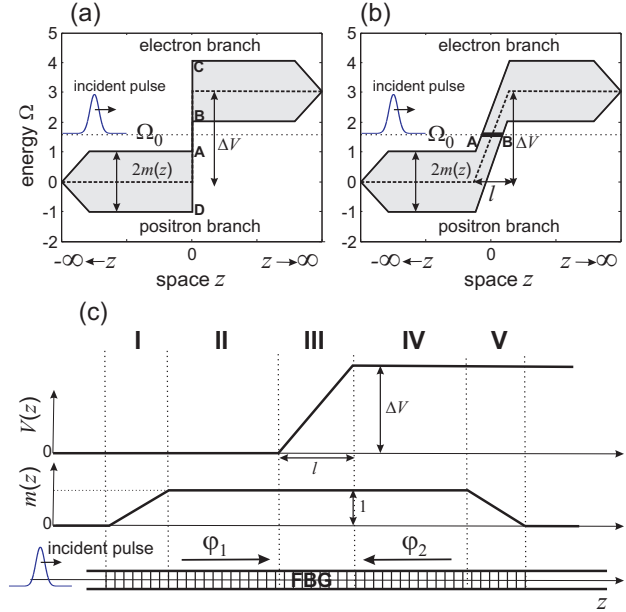}} \caption{
 Energy diagrams of the Dirac equation (3) for (a) a
sharp, and (b) a smooth potential step $V(z)$ of height $\Delta V$.
The shaded regions are the forbidden energies that separate the
electron and positron states, the dotted horizontal line is the
energy $\Omega_0$ of the incoming wave packet, and the dashed curve
is the shape of the potential step $V(z)$. (c) Schematic of the
grating structure that realizes the optical analogue of relativistic
tunneling across a potential step. The grating comprises five
sections, denoted by roman numbers (from ${\rm I}$ to ${\rm V}$),
that are defined by the amplitude $m(z)$ and phase gradient $V(z)=(d
\phi/dz)$ grating profiles.}
\end{figure}
In its present form, Eq.(3) is formally analogous to the
one-dimensional Dirac equation with $\hbar=c=1$ in presence of an
external electrostatic potential $V(z)$,  $m$ playing the role of a
dimensionless (and generally space-dependent) rest mass (see, for
instance, \cite{Calogeracos99,Emilio}). As is well-known, a
nonvanishing mass $m$ is responsible for the existence of a
forbidden energy region, which separates the positive- and
negative-energy branches of the massive Dirac equation. The optical
analogue of the forbidden energy region is precisely the photonic
stop band of the periodic grating. As the refractive index
modulation of the grating, i.e. the mass term $m$ in the Dirac
equation (3), is decreased, the stop band region shrinks and the
limit of a massless Dirac equation (similar to the one describing
the dynamics of electrons in graphene near a Dirac point) is
attained. The additional external potential $V$ in Eq.(3), related
to the chirp of the grating according to $V(z)=(d \phi/dz)$, changes
the local position of the forbidden energy region.  Therefore, pulse
propagation in a FBG with a suitably designed chirp profile can be
used to mimic the relativistic tunneling of a wave packet in a
potential step $V(z)$. It should be noted that, as compared to other
photonic analogues of KT  recently proposed in
Refs.\cite{Segevun,Longhi10} and based on {\it spatial} light
propagation in periodic photonic structures, the phenomenon of KT
occurring in FBGs and discussed in the following section involves
the {\it temporal} (rather than the spatial) light dynamics and can
be therefore simply investigated in the frequency domain by
spectrally-resolved transmission measurements.

\section{Klein tunneling}
To realize the analogue of KT, let us first assume that the optical
pulse propagates in a region of the grating where $m(z)$ is uniform
and equal to one, and let us assume a chirp profile that mimics a
step potential with an increase from $V=0$ to $V=\Delta V$ which
occurs over a length $l$ (see Fig.1). Since for the Dirac equation
(3) written in dimensionless units the Compton length is
$\lambda_C=1$ and the rest energy is $mc^2=1$, according to Sauter's
analysis KT is expected to be observable for $l$ smaller than $\sim
1$ and for a potential height $\Delta V$ larger than $2$
\cite{Sauter,Calogeracos99,Emilio}. The process of KT and tunneling
inhibition for a smooth potential step can be simply explained by a
graphical analysis of the space-energy diagrams $(z,\Omega)$ of the
one-dimensional Dirac equation \cite{Calogeracos99}, which are shown
in Figs.1(a) and (b) for a sharp and for a smooth potential step,
respectively. For the sake of clearness, in the figures the
potential $V(z)$ has been chosen to yield a nonvanishing and
constant chirp rate over a length $l$; different forms for the
potential step, such as the profile $V(z)=(\Delta V/2)[1+{\rm
tanh}(z/l)]$ considered in the seminal work by Sauter \cite{Sauter},
can be assumed as well without changing the main results.
\begin{figure}[htb]
\centerline{\includegraphics[width=8.2cm]{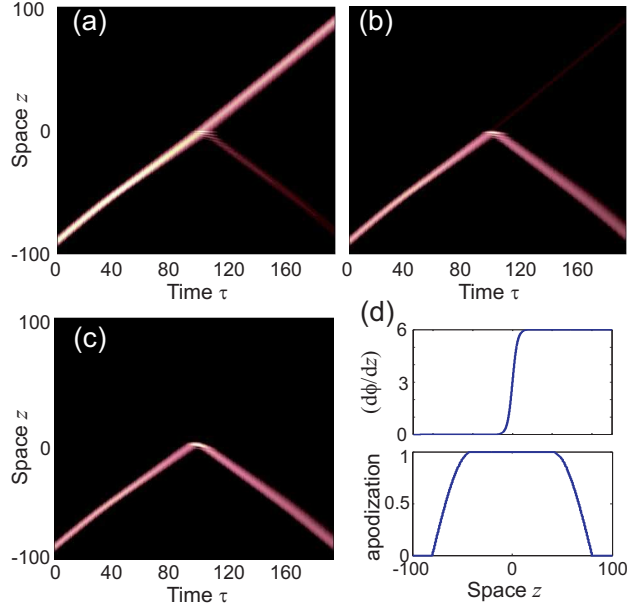}} \caption{ (a-c)
Pulse propagation in a FBG with a chirp profile $V(z)=(\Delta V /2)
[1+{\rm tanh} (z/l)]$ for $\Delta V=6$ and for: (a) $l=0.1$, (b)
$l=1.5$, and (c) $l=5$. (d) Profiles of $V=(d \phi/ dz)$ (upper
plot) and of grating amplitude $m(z)$ (apodization profile, lower
plot). The potential $V$ is shown for $l=5$, corresponding to the
simulation of Fig.2(c).}
\end{figure}
The space-energy diagrams of Figs.1(a) and (b) schematically show
the behavior of the energy spectrum of Eq.(3) versus $z$, which is
composed by two branches -the electron and positron energy branches
of the Dirac equation- separated by a gap of width $2 m(z)$ and
centered along the curve $\Omega=V(z)$. The gap regions are
visualized in the diagrams by the shaded areas. A wave packet
(optical pulse) in the electron branch with an initial mean energy
$\Omega_0$ ($1<\Omega_0<\Delta V-1$) coming from $z \rightarrow
-\infty$ tunnels into the $z>0$ region after crossing a forbidden
energy region, indicated by the bold segment AB in Fig.1(b), which
vanishes for a sharp potential step [$l=0$, see Fig.1(a)]. According
to Sauter's analysis \cite{Sauter,Calogeracos99}, the tunneling
probability is appreciable provided that $l$ is smaller than $\sim
1$. In the FBG context, the energy diagrams of Fig.1 are equivalent
to the band-reflection diagrams introduced by Poladian for a
graphical analysis of nonuniform gratings \cite{Poladian}, where the
energy $\Omega$ represents the frequency detuning of the incoming
wave from the Bragg frequency $\omega_B$. The Sauter's condition $l<
\sim 1$ for KT can be derived following the analysis of
Ref.\cite{Poladian} by computation of the transmittance of the
effective grating associated to the evanescent region AB shown in
Fig.1(b) (see Sec.V.A of Ref.\cite{Poladian}). In the previous
discussion, we assumed $m(z)=1$, however for a grating with finite
spatial extent one has $m(z) \rightarrow 0$ as $z \rightarrow \pm
\infty$. To inject and to eject the optical pulse into the $m(z)=1$
grating region around $z=0$, an input and an output apodization
sections can be introduced, which adiabatically convert the input
and output wave packets from the $m(z)=0$ regions into the $m(z)=1$
grating region (see Fig.1). Therefore, the general structure of the
FBG that realizes a photonic analogue of relativistic tunneling
across a potential step consists of five sections, as shown in
Fig.1(c): two boundary apodization sections (regions I and V), and
two uniform sections (regions II and IV) separated by a central
chirped section of length $\sim l$ (region III). In Figs. 2(a-c)
typical examples of pulse tunneling across the potential step
$V(z)=(\Delta V /2)[1+{\rm tanh}(z/l)]$ are presented, showing KT
for a sharp potential step [Fig.2(a)] and inhibition of tunneling as
the step gets smooth [Fig.2(b) and (c)]. The figures depict the
temporal evolution of
$|\psi_1|^2+|\psi_2|^2=|\varphi_1|^2+|\varphi_2|^2$ -which is
proportional to the field intensity averaged in time over a few
optical cycles and in space over a few wavelengths- as obtained by
numerical analysis of Eqs.(1) and (2) for a grating length of
$z=160$ with a quarter-cosine apodization profile [see Fig.2(d)],
$\Delta V=6$, and for a few values of $l$. A forward-propagating
Gaussian pulse $\varphi_1$ of mean energy $\Omega_0=2$ coming from
$z \rightarrow -\infty$ and of duration (FWHM in intensity)
$\tau_p=5$ has been assumed as an initial condition. For typical
parameter values $n_0=1.45$, $\Delta n=3.3 \times 10^{-4}$ and
$\lambda_B \equiv 2 n_0 \Lambda=1560$ nm, which apply to FBGs used
in optical communications, the spatial and temporal scales in Fig.2
are $Z \simeq 1.5$ mm and $T \simeq 7.3$ ps, respectively. Hence, in
physical units the grating length is $L \simeq 24$ cm, whereas the
optical analogue of the Compton length is $\lambda_C=Z \simeq 1.5$
mm. Such nonuniform FBG structures should be realizable with current
FBG technology based on UV continuous laser writing \cite{fab}. It
should be finally noticed that, as in an experiment the tracing of
pulse evolution in the grating (Fig.2) can be a nontrivial task, the
signatures of KT can be simply obtained from standard spectral
transmission measurements of the grating. In fact, for a given value
of $\Delta V>2$ and according to the band diagram of Fig.1(a), in
the KT regime a transmission window at $1<\Omega < \Delta V-1$ [the
segment AB in Fig.1(a)], embedded into the two gaps $\Delta
V-1<\Omega<\Delta V+1$ and $-1<\Omega<1$ [the segments BC and AD in
Fig.1(a)] should be observed in the transmission spectrum, the
suppression of KT for a smooth potential corresponding to the
lowering of such a transmission window. This is clearly shown in
Fig.3, where the spectral transmittance of the FBGs corresponding to
the simulations of Figs.2(a), (b) and (c) are depicted. Note that,
as the length $l$ of the chirped region is increased [from Fig.3(a)
to 3(c)], the transmission window embedded in the two adjacent gaps
disappears, which is the signature of KT inhibition.\\
\begin{figure}[htb]
\centerline{\includegraphics[width=8.2cm]{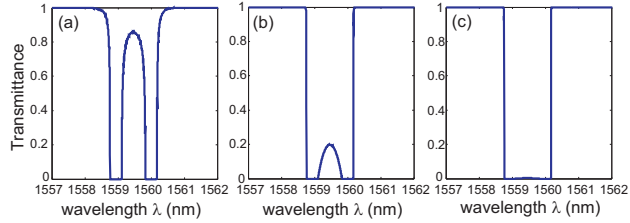}} \caption{
Numerically computed spectral transmittance of FBGs used in
numerical simulations of Fig.2 for $n_0=1.45$, $\Delta n=3.3 \times
10^{-4}$, $\lambda_B =1560$ nm, corresponding to a grating length $L
\simeq 24$ cm,  and for increasing values of $l$: (a) $l=0.1$, (b)
$l=1.5$, and (c) $l=5$.}
\end{figure}

\section{Conclusions}
In conclusion, a photonic analogue of Klein tunneling based on pulse
propagation in nonuniform fiber Bragg gratings has been proposed. As
compared to other photonic analogues of KT recently proposed in
Refs. \cite{Segevun,Longhi10} and based on spatial light propagation
in periodic photonic structures, the phenomenon of KT in FBGs
suggested in this work can be simply observed in the frequency
domain as the opening of a transmission window inside the grating
stop band, provided that the impressed chirp is realized over a
length of the order of the analogue of the Compton wavelength. Such
a FBG-based system might be thus considered to be the simplest
optical analogue proposed so far to observe KT.

\section{Acknowledgements}
The author acknowledges financial support by the Italian MIUR (Grant
No. PRIN-2008-YCAAK project "Analogie ottico-quantistiche in
strutture fotoniche a guida d'onda").

%\clearpage
%\bibliography{H:/Physik/bibliography}

\end{document}